\documentclass{article}

\usepackage{PRIMEarxiv}

\usepackage[utf8]{inputenc} 
\usepackage[T1]{fontenc}    
\usepackage{hyperref}       
\usepackage{url}            
\usepackage{booktabs}       
\usepackage{amsfonts}       
\usepackage{nicefrac}       
\usepackage{microtype}      
\usepackage{lipsum}
\usepackage{fancyhdr}       
\usepackage{graphicx}       
\graphicspath{{media/}}     

 \usepackage{array,multirow}
\usepackage{amsmath,amsfonts}
\usepackage{algorithmic}
\usepackage{array}
\usepackage[caption=false,font=normalsize,labelfont=sf,textfont=sf]{subfig}
\usepackage{textcomp}
\usepackage{stfloats}
\usepackage[ruled,vlined]{algorithm2e} 
\usepackage{verbatim}

\title{Distribution-based Low-rank Embedding
\thanks{
This is the authors' version.\\
B. Yousefi is with the University of Maryland College Park. \\

BSE 4115, 9636 Gudelsky Drive, Rockville, MD 20850, Tel: 240-665-6529, e-mail: byousefi@umd.edu}\\
}

\author{
  Bardia Yousefi
}

\begin{document}
\maketitle

\begin{abstract}
The early detection of breast abnormalities is a matter of critical significance. Notably, infrared thermography has emerged as a valuable tool in breast cancer screening and clinical breast examination (CBE). Measuring heterogeneous thermal patterns is the key to incorporating computational dynamic thermography, which can be achieved by matrix factorization techniques.
These approaches focus on extracting the predominant thermal patterns from the entire thermal sequence. Yet, the task of singling out the dominant image that effectively represents the prevailing temporal changes remains a challenging pursuit within the field of computational thermography.
In this context, we propose applying James-Stein for eigenvector (JSE) and Weibull embedding approaches, as two novel strategies in response to this challenge. The primary objective is to create a low-dimensional (LD) representation of the thermal data stream. This LD approximation serves as the foundation for extracting thermomics and training a classification model with optimized hyperparameters, for early breast cancer detection. Furthermore, we conduct a comparative analysis of various embedding adjuncts to matrix factorization methods.
The results of the proposed method indicate an enhancement in the projection of the predominant basis vector, yielding classification accuracy of 81.7\% ($\pm$5.2\%) using Weibull embedding, which outperformed other embedding approaches we proposed previously. In comparison analysis, Sparse PCT and Deep SemiNMF showed the highest accuracies having 80.9\% and 78.6\%, respectively. These findings suggest that JSE and Weibull embedding techniques substantially help preserve crucial thermal patterns as a biomarker leading to improved CBE and enabling the very early detection of breast cancer.

\end{abstract}


\section{Introduction}
Thermography is an effective method for early breast cancer detection, addressing the second leading cause of cancer death in women~\cite{r1}. This study suggests two novel methods, James-Stein for eigenvector (JSE) in thermography and the Weibull embedding, for thermography as clinical breast examination (CBE) tools and pre-screening technique before mammography, providing information on potential abnormalities in patients ~\cite{r2,r3}. 
Thermographic imaging operates on the principle of intensified angiogenesis (blood vessel formation) and vasodilation due to lesions detectable by an infrared camera. Endocrine-induced changes in the thermal profile signify altered vascularization, facilitating oxygen and nutrient delivery to lesion sites~\cite{r3,r4}. Employing an infrared camera enables the dynamic capture of these changes, enhancing the ability to detect abnormalities by measuring thermal heterogeneities. Multiple studies have confirmed the importance of thermography in identifying hypervascularity related to non-palpable breast cancer~\cite{r4}. This technique shows promise as a biomarker for early breast cancer detection, complementing CBE and preceding mammography. However, the major hurdle in using infrared technology for breast cancer detection lies in capturing irregularities and translating raw infrared sequences into thermal patterns. To overcome this, low-rank matrix approximation methods have been widely employed, where the primary basis is selected or to extract the leading basis that represents the entire thermal stream. Several computational thermographic techniques have been employed to aim this objective and used for dynamic thermography, techniques such as PCA- principal component analysis (PCA)~\cite{r5,r6}, non-negative matrix factorization (NMF)~\cite{r7}, eigenvector extraction with unchanged condition~\cite{r8}, incrementally calculated PCA~\cite{r9}, sparse matrix factorization~\cite{r10}, t-distributed stochastic neighbor embedding (t-SNE)~\cite{r11}, candid covariance-free incremental principal component thermography (CCIPCT)~\cite{r12}, sparse PCA~\cite{r13,r14}, semi-NMF~\cite{r15,r16,r17}, sparse NMF~\cite{r18}, convex NMF~\cite{r19}, deep NMF~\cite{r20}, and deep learning models integrated with convex NMF~\cite{r21}.
Low-rank matrix approximation finding with maximum thermal pattern variance requires selecting the appropriate basis, which poses an additional challenge. To address this challenge, two novel thermographic methods are proposed here:

Overall, this paper presents the following contributions:
\begin{list}{-}{}
\item{\textbf{James-Stein for eigenvector (JSE)}~\cite{r23,r24,r25} is used to estimate high-dimensional (HD) thermal sequences to low-dimensional (LD) data. JSE showed a considerable effect on the minimization of variance, which helps the correction of the predominant eigenvector for thermography.}

\item{\textbf{Weibull embedding}, which is defined by using the Weibull distribution function. Techniques such as Gaussian~\cite{r20} and Bell~\cite{r22} embedding approaches have been introduced previously, which showed significant aid in magnifying heterogeneous thermal patterns. This study demonstrates the application of Weibull embedding in the context of factorization analysis for thermography. It introduces an innovative approach that combines prominent basis elements with embedding techniques to extract critical thermomics, facilitating the training of a classifier for the early identification of breast abnormalities.}

\item{We conducted a comparative analysis of known computional thermography methods to reduce HD data, and used the latent space representation of these data to diagnose the very early breast cancer. These applications were conducted to better understand the strengths and weaknesses of the proposed approaches. Throughout all the circumstances, our embedding approaches deliver better results than the baselines.}
\end{list}
The methods proposed here have been comparatively tested on different factorization techniques and the results indicate promising diagnostic outcomes for early breast cancer detection. This approach can result in enhanced precision when depicting thermal patterns, thereby yielding a more appropriate biomarker for the intended purpose.
Table I provides an overview of the mathematical notations employed throughout this article. The subsequent section elaborates on the proposed methodology. Following that, Sections III and IV present the experimental findings and subsequent discussion, respectively. Finally, Section V encapsulates the conclusions drawn from this study.

\begin{table*}
\begin{center}
\caption{Table of notations.} \vspace{-0.0in}
\label{notations}
\includegraphics[width=0.99\linewidth]{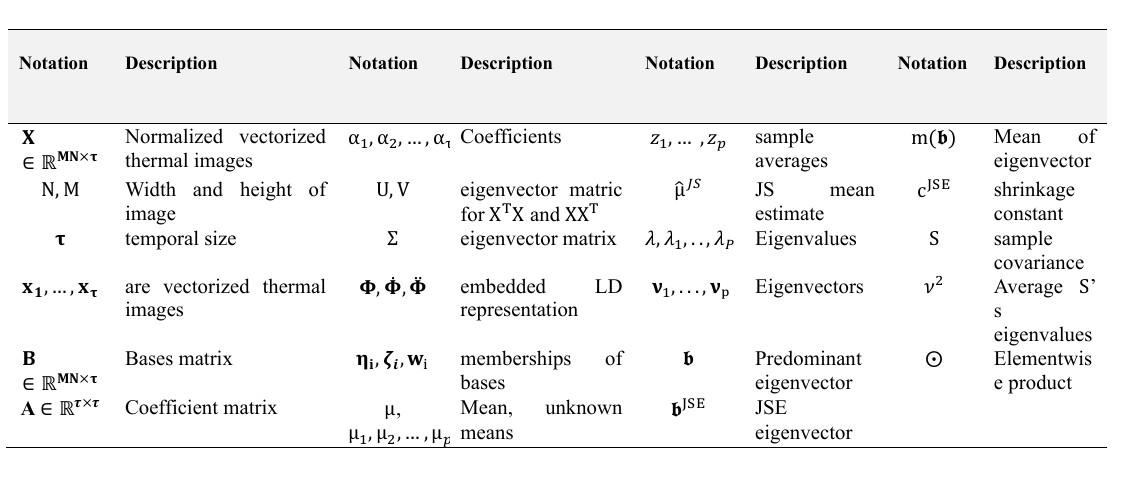} 
\end{center} \vspace{-0.3in}
\label{table-not}
\end{table*}

\begin{figure*}[t]
\begin{center}
    \includegraphics[width=0.95\linewidth]{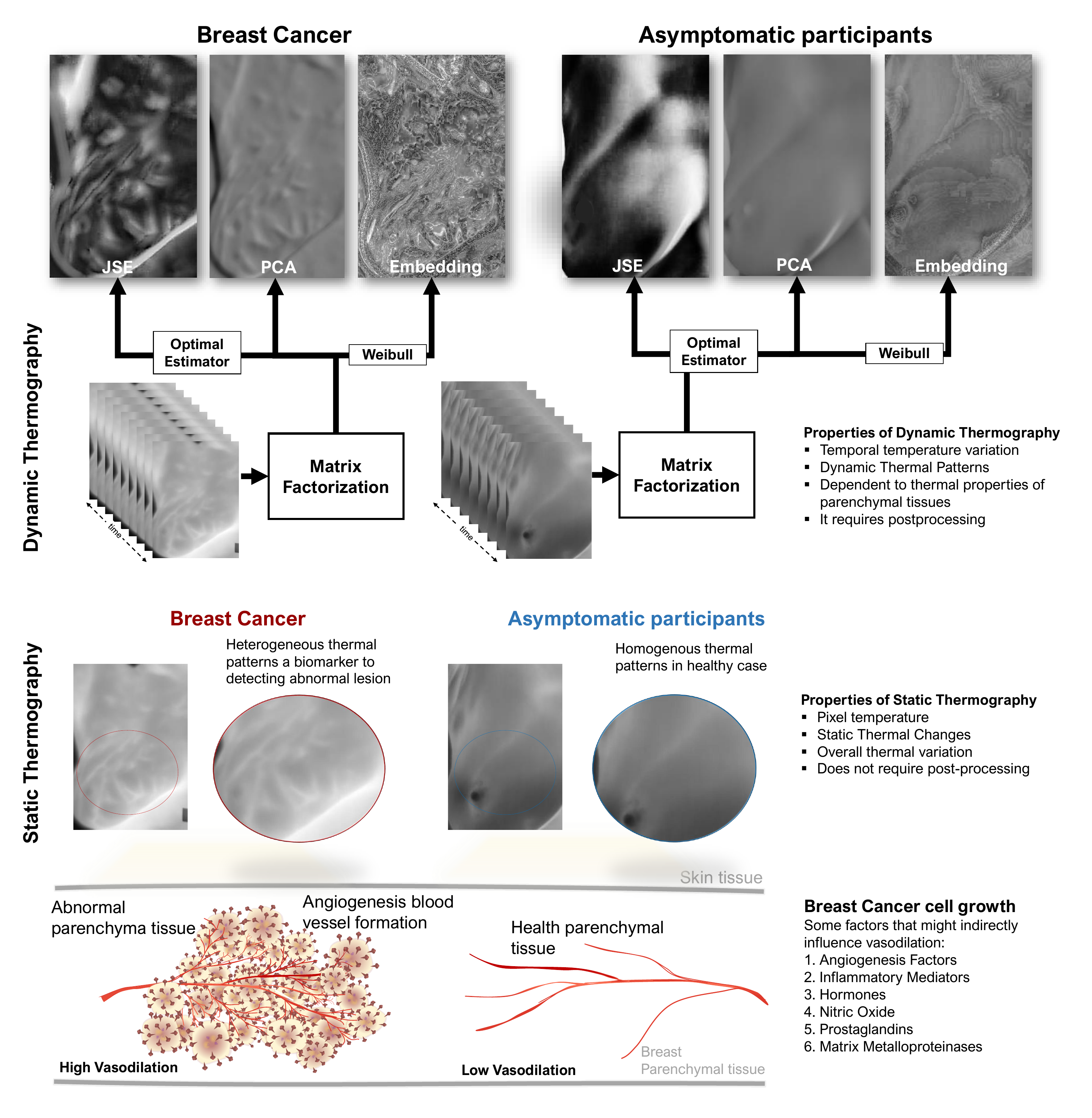}
\end{center} \vspace{-.2in}
   \caption{Scheme of thermography system setup for early detection of breast cancer leading abnormality. 
   Breast cancer exerts a multifaceted influence on vasodilation, intricately linked to various factors. Angiogenesis factors, inflammatory mediators, hormones, nitric oxide, prostaglandins, and matrix metalloproteinases collectively orchestrate a complex interplay that modulates vascular responses, underscoring the intricate relationship between breast cancer and the dynamic regulation of blood vessel function in its microenvironment.
   Dynamic thermography has significant advantage over static thermography due to increase the chance of capturing temporal thermal fluctuations which can be used as a biomarker for very early diagnosis of breast cancer in the CBE stage.}
\label{fig1}
\end{figure*}

\section{Related works and Motivations} \vspace{-.01in}
Heterogeneous thermal patterns signify irregular thermal characteristics within parenchymal tissues~\cite{r3,r4,r11,r18,r19,r20,r21}. Measuring these patterns and highlighting them can be computed by eigendecomposition and matrix factorization techniques tailored for extracting the dominant basis across thermal dimensions. These methods enable the representation of temporal pattern variations observed in the infrared images~\cite{r5,r6,r7,r8,r9,r10,r11,r12,r13,r14,r15,r16,r17,r18,r19,r20}. PCA in thermography (PCT)~\cite{r5,r6} provides a low-dimensional approximation of the thermal stream by decomposing the heat matrix through singular value decomposition (SVD)~\cite{r5,r6}. The PCT approach has been modified into CCIPCT~\cite{r9,r12} and sparse PCT~\cite{r13,r14} techniques. CCIPCT and Sparse PCT introduce incremental computation to enhance speed and incorporate additional regularization parameters to robustize PCT, respectively.
NMF can be obtained by imposing non-negative constraints on both the bases and coefficients of PCA, as discussed in prior works~\cite{r7,r16,r18}. Relaxing these constraints leads to modified NMF variants, including semiNMF~\cite{r17,r19} and Convex-NMF~\cite{r18}. Sparse NMF~\cite{r17,r18,r19}, similar to Sparse PCT~\cite{r13,r14}, incorporates regularization terms to strengthen the decomposition process, establishing methodological comparability. The deep semi-NMF algorithm decomposes the basis matrix into multiple hidden bases, ensuring a sparse representation. Deep layers of bases can be trained to effectively preserve diverse thermal patterns~\cite{r20}.
These methods rely on pairwise distances between temporal points within thermal sequences to maintain the relative distances when transforming them into a lower-dimensional space. Applying t-SNE replaces the typical Euclidean distance between pairwise points within thermal sequences with a probabilistic similarity measure. By minimizing the Kullback-Leibler divergence between probability distributions of two-point centered Gaussian distributions, this approach enhances the data representation~\cite{r11}.  Selecting the most prominent bases is a notable challenge, which we tackle by utilizing JSE and Weibull embedding techniques and conducting an extensive comparative analysis of existing models, ensuring the robustness and reliability of the diagnostic system.


\section{Method} \vspace{-0.05in}
\noindent Measuring heterogeneous thermal patterns in infrared imaging is critical for early breast cancer diagnosis. Here, we put forward the application of the JSE and Weibull embedding methods for the detection of thermal patterns via factorization analysis. In addition, we introduce a novel embedding technique in an infrared-based diagnostic system (see Fig. 1 and Fig.2).

\subsection{Overview of Low-rank representation thermography }
The input data X consists of a stack of vectorized thermal images, forming a heat matrix. A low-rank approximation model can be expressed as follows:
\begin{equation}
    X  \approx BA
\end{equation}
where $X \in \mathbb{R}^{MN \times \tau}$, \textit{i.e.}, $X=[\mathbf{x}_1,\mathbf{x}_2,\dots,\mathbf{x}_{\tau} ]$ and can be shown by a linear combination of $\tau$ basis vectors, $B= \{ \mathbf{\beta}_1,\mathbf{\beta}_2,\dots,\mathbf{\beta}_\tau \}, B \in \mathbb{R}^{MN \times \tau}$ and $A$ coefficient matrix, $A \in \mathbb{R}^{\tau \times \tau}, A = \{ \mathbf{\alpha}_1,\mathbf{\alpha}_2,\dots,\mathbf{\alpha}_\tau \}$. $\mathbf{x}_1,\mathbf{x}_2,\dots,\mathbf{x}_\tau$ are vectorized thermal frames. The matrix $X$ represents normalized stacked thermal tensors derived from input thermal images. In PCT, equation (1) is presented as:
\begin{equation}
    X= U \Sigma V^T
\end{equation}
where $\Sigma$ denotes as singular values of $X$. $U$ and $V$ are eigenvector matrices for $XX^T$ and $X^TX$, respectively. Basis vectors and coefficients can be modified depending on which eigendecomposition is used. For example, in PCT~\cite{r13,r14}, CCIPCT~\cite{r12}, and SparsePCT~\cite{r13,r14} can be negative and non-negative, whereas in NMF~\cite{r7,r16,r18}, semiNMF~\cite{r17,r19}, convex NMF~\cite{r18} there are non-negative constraints restraining coefficients or base.
Here we use JSE as a corrector for estimating the leading eigenvector in eigendecomposition.

\begin{figure*}[t]
\begin{center}
   \includegraphics[width=0.95\linewidth]{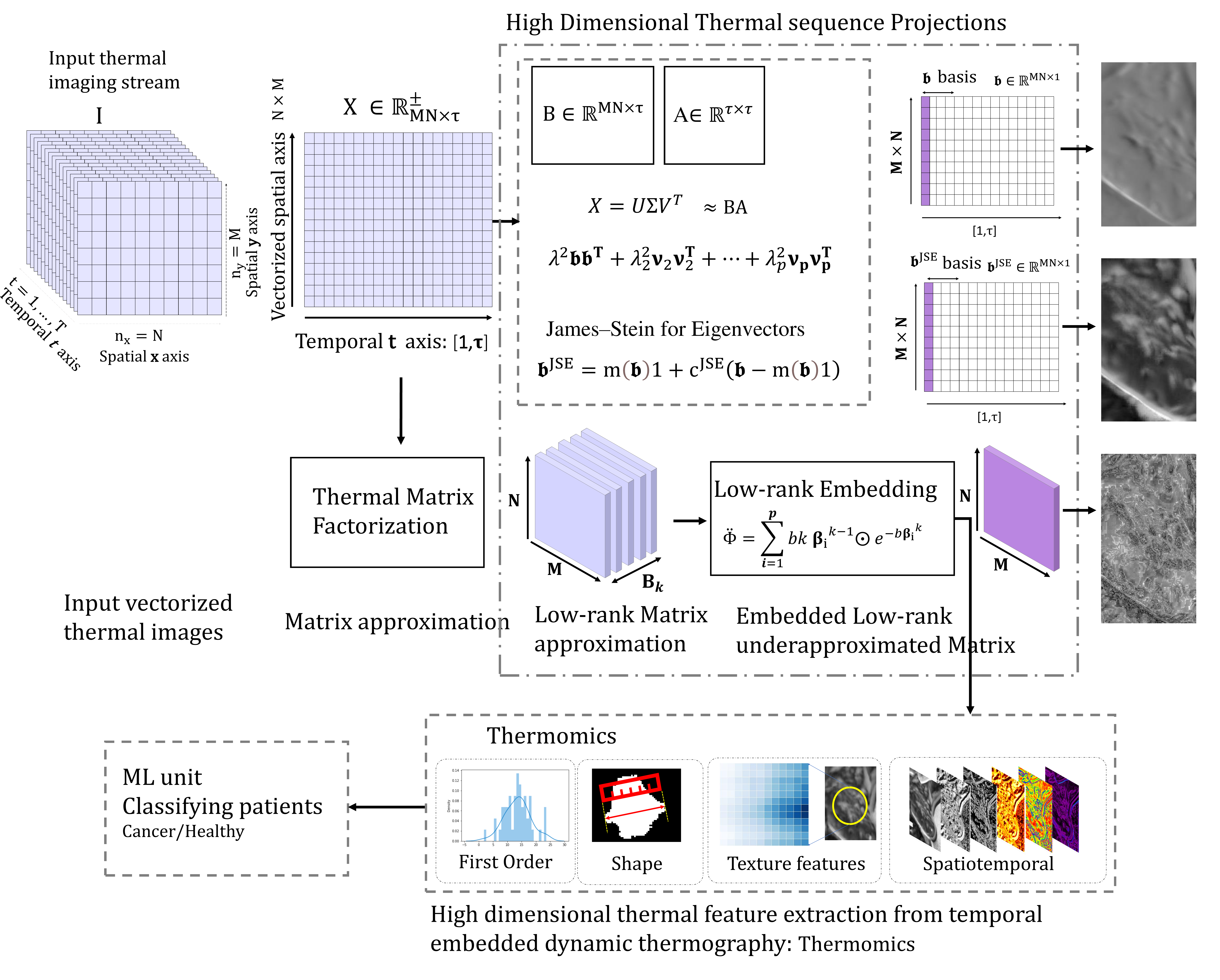}
\end{center}
   \caption{Workflow of applying JSE with low-rank Weibull embedding in thermography.}
\label{fig2}
\end{figure*}

\subsection{Low-rank Embedding}
The use of basis vector embedding has been proposed in previous studies~\cite{r20,r22}, which involves combining multiple decomposed vectors to decrease the dimensionality of the heat matrix of stacked infrared images. We previously applied low-rank representation methods to transform higher temporal dimensionality into lower temporal representations and can be regarded as basis vectors that are computed using matrix factorization techniques. We generate a set of LD projected basis vectors ($B$) and integrate their overall representation using two embedding membership functions~\cite{r20,r22}. This approach allows us to effectively perform dimensionality reduction of presenting thermal images while preserving important information. From~\cite{r20,r22}, we adopted:

\newtheorem{definition}{\textbf{Definition}}
\begin{definition}
\noindent { \it The embedded low-dimensional representation, $\mathbf{\Phi}$, defines by aggregating membership calculated for $p$ basis vectors of $X$, $\mu_p$, elementwise multiply by the basis vector itself, $\mathbf{\beta_i}$, and defined as:}
    \begin{equation}
       {\LARGE \mathbf{\Phi} = \sum_{i=1}^p \mathbf{\beta}_i \odot \mathbf{\eta}_i }
    \end{equation}
 {\it where $\mathbf{\eta}_i$ is a membership on a basis vector $\mathbf{\beta}_i$ and defines by: }\\
 \begin{equation*}
     { \LARGE \mathbf{\eta}_i = e^\frac{\mathbf{\beta}_i-\mu}{\sigma}}.
 \end{equation*}
\end{definition}

\noindent Let $\mu$,$\sigma$ mean (average) of a thermal image in the region of interest (ROI) basis and standard deviation of $i^th$-vector in the calculation. In this definition $\Phi \in \mathbb{R} ^{s \times 1}$, and $X \in \mathbb{R} ^{s \times \tau}$  and $p \ll \tau$. \\
A derivation of this definition, while we use Cauchy distribution, is defined as follows:

\begin{definition}
\noindent { \it The embedded low-dimensional representation, $\mathbf{\dot{\Phi}}$, defines by aggregating membership calculated for $p$ basis vectors of $X$, $\mu_p$, elementwise multiply by the basis vector itself, $\mathbf{\beta_i}$, and defined as:}
    \begin{equation}
       {\LARGE \mathbf{\dot{\Phi}} = \sum_{i=1}^p \mathbf{\beta}_i \odot \mathbf{\xi}_i }
    \end{equation}
 {\it where $\mathbf{\xi}_i$ is a membership on a basis vector $\mathbf{\beta}_i$ and defines by: }\\
 \begin{equation*}
     { \LARGE \mathbf{\xi}_i = \frac{1}{1+|\frac{\mathbf{\beta}_i - \mu}{\sigma}|^2b}}.
 \end{equation*}
\end{definition}

\noindent we use this embedding with an obituary coefficient $b$ (set to 1) and a generalized bell curve $\mathbf{\xi}_i$, which is a variation of the Cauchy distribution, to combine the LD-represented bases into an overall representation. 
Both embedding approaches emphasize thermal pattern variation, leading to exponentially enhanced heterogeneity in the resulting image.

\subsection{Weibull Embedding}
The introduction of Weibull embedding involves the representation of the probability density related to a Weibull random variable~\cite{r26,r27,r34,r35}, as follows:
\begin{equation}
f(x;\lambda ,k) = \ \left\{ \begin{matrix}
\frac{k}{\lambda}\left( \frac{x}{\lambda} \right)^{k - 1}e^{- \left( \frac{x}{\lambda} \right)^{k}} & x \geq 0 \\
0, & x < 0 \\
\end{matrix} \right.\ \ \ \text{\ }\ 
\end{equation}

The probability density function of a Weibull random variable is characterized by two parameters: the shape parameter ($k\  > \ 0$) and the scale parameter ($\lambda\  > \ 0$). The complementary cumulative distribution function of the Weibull distribution follows a stretched exponential function. This distribution shows connections with various other probability distributions and notably acts as an intermediary between the exponential distribution (k = 1) and the Rayleigh distribution ($k = 2$ and $\lambda = \sqrt{2\sigma}$~\cite{r28}).\\

\begin{definition}
\noindent { \it The embedded low-dimensional representation, $\mathbf{\ddot{\Phi}}$, defines by aggregating membership calculated for $p$ basis vectors of $X$, $\mu_p$, elementwise multiply by the basis vector itself, $\mathbf{\beta_i}$, and defined as:}
    \begin{equation}
       \ddot{\mathbf{\Phi}} = \sum_{i = 1}^{p}{\mathbf{w}_{i}\ }
    \end{equation}
 \emph{where} \(\mathbf{w}_{\mathbf{i}}\) \emph{is a function of a basis
vector} \(\mathbf{\beta}_{i}\) \emph{and defines by for element of}
\(\mathbf{\beta}_{i}^{e} \geq 0\): \\
 \begin{equation*}
     \mathbf{w}_{\mathbf{i}}\mathbf{=}bk\ {\mathbf{\beta}_{i}}^{k - 1}{\odot\ e}^{- b{\mathbf{\beta}_{i}}^{k}}
 \end{equation*}
{ \it where the shape parameter \(k\) is the same as above, while the scale
parameter is \(b = \lambda^{- k}.\)}
\end{definition}

\subsection{James--Stein for Eigenvectors Thermography}

Let's consider a scenario in which the number of unknown means, denoted as $\mu= (\mu_{1},\mu_{2},\dots,\mu_{p})$, where $p>3$, exceeds three, and the objective is to estimate these
means. To achieve this, we observe a pre-determined number of samples and calculate their respective sample means, denoted as $z = (z_{1},z_{2},\dots\, z_{p})$. When estimating the unobserved mean value $\mu_{i}$, it is often convenient to employ the sample averages $z_{i}$ as estimates. This approach is particularly effective when approximating a single mean. Seminal work by Stein~\cite{r29} and James and Stein~\cite{r30} has demonstrated that a superior estimator can be achieved by shrinking the sample means towards their collective mean. James and Stein demonstrated that for $p>3$, the James-Stein estimator (${\widehat{\mu}}^{JS}$) outperforms $z$ in terms of expected mean squared error.
\begin{equation}
   E_{\mu,\nu} \left[\left| \widehat{\mu}^{JS} - \mu \right|^{2} \right] < E_{\mu,\nu} \left[\left| z - \mu \right|^{2} \right]. 
\end{equation}
\noindent where ${\mu\mathbb{\in R}}^{p},\ \nu\mathbb{\in R.\ }\)An extension of the James-Stein estimator involves applying it to eigenvectors. The general eigen decomposition problem derived from equation (2), especially in the context of PCA, entails minimizing variance and fluctuations in thermal patterns, as follows:
\begin{equation}
\begin{aligned}
& & \min_{U \in \mathbb{R}^{p}}{U^{T}\Sigma U}   \\
& & U^{T}\mathbf{1}_{p} = 1.
\end{aligned}               
\end{equation}

\noindent We estimate $\Sigma$ and obtain $U$. This involves considering a sequence of n independent observations of a p-dimensional variable, where the population's covariance matrix, denoted by $\Sigma$, is unknown. We can decompose the sample covariance matrix $S$ with $p \times p$ dimensions spectrally as follows:
\begin{equation*}
S = \lambda^{2}\mathfrak{b}\mathfrak{b}^{\mathbf{T}} + \lambda_{2}^{2}\mathbf{\nu}_{\mathbf{2}}\mathbf{\nu}_{\mathbf{2}}^{\mathbf{T}} + \lambda_{3}^{2}\mathbf{\nu}_{\mathbf{3}}\mathbf{\nu}_{\mathbf{3}}^{\mathbf{T}} + \ldots + \lambda_{p}^{2}\mathbf{\nu}_{\mathbf{p}}\mathbf{\nu}_{\mathbf{p}}^{\mathbf{T}}.
\end{equation*}

\noindent The decomposition of the sample covariance matrix $S$ results in non-negative eigenvalues $\lambda^{2} \geq \lambda^{12} \geq \lambda^{22} \geq \dots \geq \lambda ^{2} \geq 0$
with their corresponding orthonormal bases $\left\{ \mathfrak{b}, \mathbf{\nu}^{\mathbf{2}}\mathbf{,\ \ldots,\ \nu } \right\}$ of $S$. In scenarios where $p$ significantly exceeds n, our main focus centers on the leading eigenvalue $\lambda_{1}$ and its corresponding eigenvector $\mathfrak{b}$. It is important to highlight that, within the context of the preceding discussion, the sample eigenvector $\mathfrak{b}$ serves a role akin to the set of sample averages $z$.

\noindent In classical statistics, when the population eigenvalues are distinct and $p$ is constant, the sample eigenvalues and eigenvectors act as dependable estimators of their respective population counterparts. However, this characteristic might not hold true as the dimension $p$ approaches infinity. Within the high to low-dimensional projection, also known as the HL setting, the JSE method seeks to provide an empirical estimator that enhances the sample eigenvector $\mathfrak{b}$.

JSE, similar to the JS estimator, serves as a shrinkage estimator tailored for improving the performance of the sample eigenvector $\mathfrak{b}$ in the HL setting. Although sample eigenvalues and eigenvectors are trustworthy estimators of their population counterparts, JSE focuses on reducing the squared error of the eigenvector $\mathfrak{b}$ with high probability, thereby enhancing estimation accuracy. This improved accuracy is particularly valuable for obtaining precise covariance matrix estimates in quadratic optimization tasks.

The JSE estimator, denoted by $\mathfrak{b}^{JSE}$, is obtained by shrinking the elements of the sample eigenvector $h$ towards their mean value $m(\mathfrak{b})$, similar to the JS approach~\cite{r23}. 
\[\mathfrak{b}^{JSE} = m\left( \mathfrak{b} \right)\mathbf{1} + c^{JSE}(\mathfrak{b} - m\left( \mathfrak{b} \right)\mathbf{1})\]
\noindent where the shrinkage constant $c^{JSE}$ is
\[c^{JSE} = 1 - \frac{\nu^{2}}{s^{2}\left( \mathfrak{b} \right)},\]
\noindent where
\[s^{2}(h) = \frac{1}{p}\sum_{i = 1\ }^{p}\left( \lambda\mathfrak{b}_{i} - \lambda m(\mathfrak{b}) \right)^{2}\]
The quantity $s^{2}$ measures the variability of the entries of $\lambda\mathfrak{b}$ around their mean value $\lambda m(\mathfrak{b})$, while $\nu^{2}$ is the mean of the smaller nonzero $S$ eigenvalues, multiplied by $1/p$,
\[\nu^{2} = \frac{tr(S) - \lambda^{2}}{p.(n - 1)}\ .\]
\noindent Similar to the James-Stein estimator, the JSE also relies on shrinkage towards a central estimate. This shrinkage is more pronounced when the smaller, non-zero, eigenvalues of the $S$ dominate the variation of the leading eigenvector, and less pronounced otherwise. In other words, S controls the $\lambda\mathfrak{b}$ variation about their mean. This reduces the angular distance between the  sample leading eigenvector and the population eigenvector, and the JSE improves upon the estimation of
the leading eigenvector $\mathfrak{b}$ of $S$. Factor models are commonly used to reduce the dimensionality of complex outcomes by identifying a smaller set of driving factors~\cite{r10,r11}.

\subsection{Extracting thermomics, imaging biomarkers}
Extracting quantitative attributes from diverse medical imaging data within a region of interest (ROI) is instrumental for enhancing computer-aided decision-making in cancer imaging analyses. This process involves the utilization of various filters to process and interpret information, generating analytical responses crucial for diagnosis. 
Within the infrared system, these attributes are termed thermomics~\cite{r17,r18,r19,r20}, extensively employed in the early detection of breast cancer. Termed heterogeneous thermomics, these attributes serve as
biomarkers or indicators, highlighting abnormal thermal patterns within breast tissues~\cite{r15,r17,r18,r19,r20,r21,r22,r23}. High-dimensional (HD) thermomics are derived from the predominant basis using the Pyradiomics library~\cite{r24}.

To address the concern of overfitting in the decision-making model, we applied spectral embedding to reduce the dimensionality of the thermomics, minimizing the risk. Subsequently, we conducted statistical analyses to establish connections between pairwise thermal features. 
Here, we introduce two approaches designed to enhance early breast cancer diagnosis, leveraging convex factorization embedding within thermal analysis. Particularly, we utilize the Convex-NMF method to derive thermomics from a series of dynamic thermographic images. To improve the representation of the extracted bases, we introduce a novel embedding technique known as Weibull embedding, providing an alternative to the conventional Bell and Gaussian embedding techniques. This innovative approach seeks to enhance the precision and dependability of the diagnostic system by focusing on pertinent thermal patterns within the breast tissue.

\begin{table}[]
\begin{center}
    
\caption{Demographics and clinical information of thermal breast cancer screening database.
}
\begin{tabular}{lll}
\hline
\multicolumn{3}{c}{\textbf{DMR - Database for   Mastology Research}}                                                                                                                                                                  \\ \hline
Age                    & Median (±IQR)                                                                                           & 60 (25,120)                                                                                                       \\
Race                   & \begin{tabular}[c]{@{}l@{}}Caucasian\\    African\\    Pardo\\    Mulatto \\    Indigenous\end{tabular} & \begin{tabular}[c]{@{}l@{}}78 (36.8\%)\\    57 (26.9\%)\\    75 (35.4\%)\\    1 (0.5\%)\\    1 (0.5\%)\end{tabular} \\
Diagnosis$^1$             & \begin{tabular}[c]{@{}l@{}}Healthy$^2$\\    \\ Symptomatic  \\ Sick$^3$\end{tabular}                            & \begin{tabular}[c]{@{}l@{}}131 (61.7\%) \\    \\ 81 (38.2\%)\\    36 (16.9\%)\end{tabular}                        \\
Family   history       & \begin{tabular}[c]{@{}l@{}}Diabetes\\    Hypertensive\\    Leukemia\\    None\end{tabular}              & \begin{tabular}[c]{@{}l@{}}52 (24.5\%)\\    5 (2.3\%)\\    1 (0.5\%)\\    154 (72.6\%)\end{tabular}                 \\
Hormone   therapy (HT) & \begin{tabular}[c]{@{}l@{}}Hormone   replacement\\    None\end{tabular}                                 & \begin{tabular}[c]{@{}l@{}}38 (17.9\%)\\    174 (82.1\%)\end{tabular}                                             \\ \hline
\end{tabular} \\
\end{center}
{\footnotesize
1. The assessment carried out utilizing mammography as the ultimate verity within this dataset. 2. The designation "healthy" pertains to individuals devoid of malignancy and lacking any evident symptoms. 3. We employ the terminology "sick" to encompass an array of breast cancer sufferers identified via mammographic imaging.

}
\label{table1}

\end{table}

\section{Results}
The monitoring of vasodilation and angiogenesis, key factors in breast cancer, was achieved through the analysis of a range of thermal patterns in breast cancer screening datasets~\cite{r31}. To provide a comprehensive insight, we utilized convex factorization embedding. Moreover, a comparative analysis was conducted, juxtaposing the results from convex factorization embedding against those derived from various low-rank matrix approximation techniques. This meticulous examination aimed to elucidate the nuances and discrepancies inherent in these methods, fostering a deeper understanding of their relative effectiveness.
\subsection{Study Population}
A total of 212 participants, encompassing both individuals without any symptoms (healthy) and those afflicted with various conditions (including symptomatic cases and cancer patients), were carefully selected to establish a robust benchmark for the proposed model. For this study, cancer patients and symptomatic cases underwent clinical breast examination (CBE), mammography imaging, and tissue biopsy, while cancer patients and patients with symptoms were categorized as abnormal cases. The study cohort exhibited a median age of 60 years, comprising 78 individuals of Caucasian ethnicity (36.8$\%$), 57 individuals of African descent (26.9$\%$), 75 individuals of Pardo origin (35.4$\%$), 1 indigenous individual (0.5$\%$), and 1 individual of Mulatto heritage (0.5$\%$). Among the participants, 38 individuals (17.9$\%$) were undertaking hormone replacement therapy, and 52 cases (25$\%$) had a family history of diabetes. \\
To capture comprehensive thermal information, all patients underwent infrared acquisition following a standardized protocol, with images possessing a spatial resolution of $640 \times 480$ pixels. The FLIR thermal camera (model SC620) was employed, featuring a sensitivity within the range of less than $0.04^{\circ}C$ and capable of capturing a broad thermal range spanning from $-40^{\circ}C$ to $500^{\circ}C$~\cite{r31}. Further details regarding the demographics and clinical information of the study cohort are presented in Table II.

\subsection{Results of JSE in thermography}
\begin{figure}[t]
\begin{center}
   \includegraphics[width=0.85\linewidth]{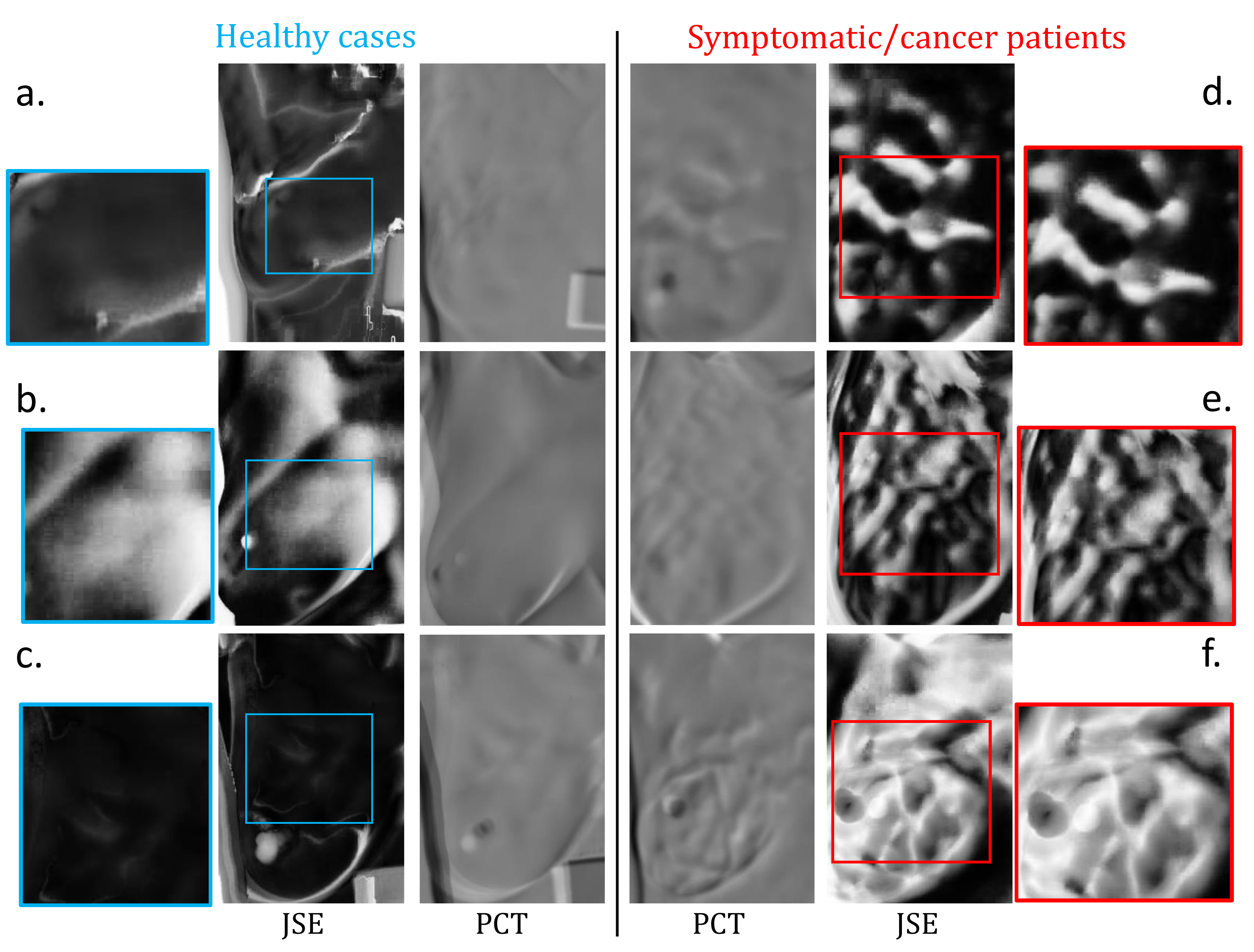}
\end{center}
   \caption{Six examples of PCT versus JSE thermography results, three healthy cases (a-c) and three abnormal (d-f) cases. JSE unequivocally bolsters heterogeneous thermal pattern underlining for vasodilation in breast cancer cases. }
\label{fig3}
\end{figure}

PCT was applied to project the 23-dimensional thermal stream onto 5 low-dimensional representative basis vectors, followed by the use of the JSE estimator. Figure \ref{fig3} provide illustrative examples of the resulting low-dimensional (LD) representations from PCT and JSE. These LD images effectively capture the diverse and heterogeneous textures found in the regions of interest (ROI) in over 80 cases during breast cancer screening, including both individuals with symptoms and healthy participants. In contrast, the LD approximation of thermal variations in the healthy participants exhibits a higher degree of homogeneity, as shown in Figure \ref{fig3} (a-c). The cross-validated accuracy of binary classification using JSE yield 76.2\% (71.4\%,80.9\%) using random forest tuned classifier.

\subsection{Results of Weibull embedding}
\begin{figure}[t]
\begin{center}
   \includegraphics[width=0.85\linewidth]{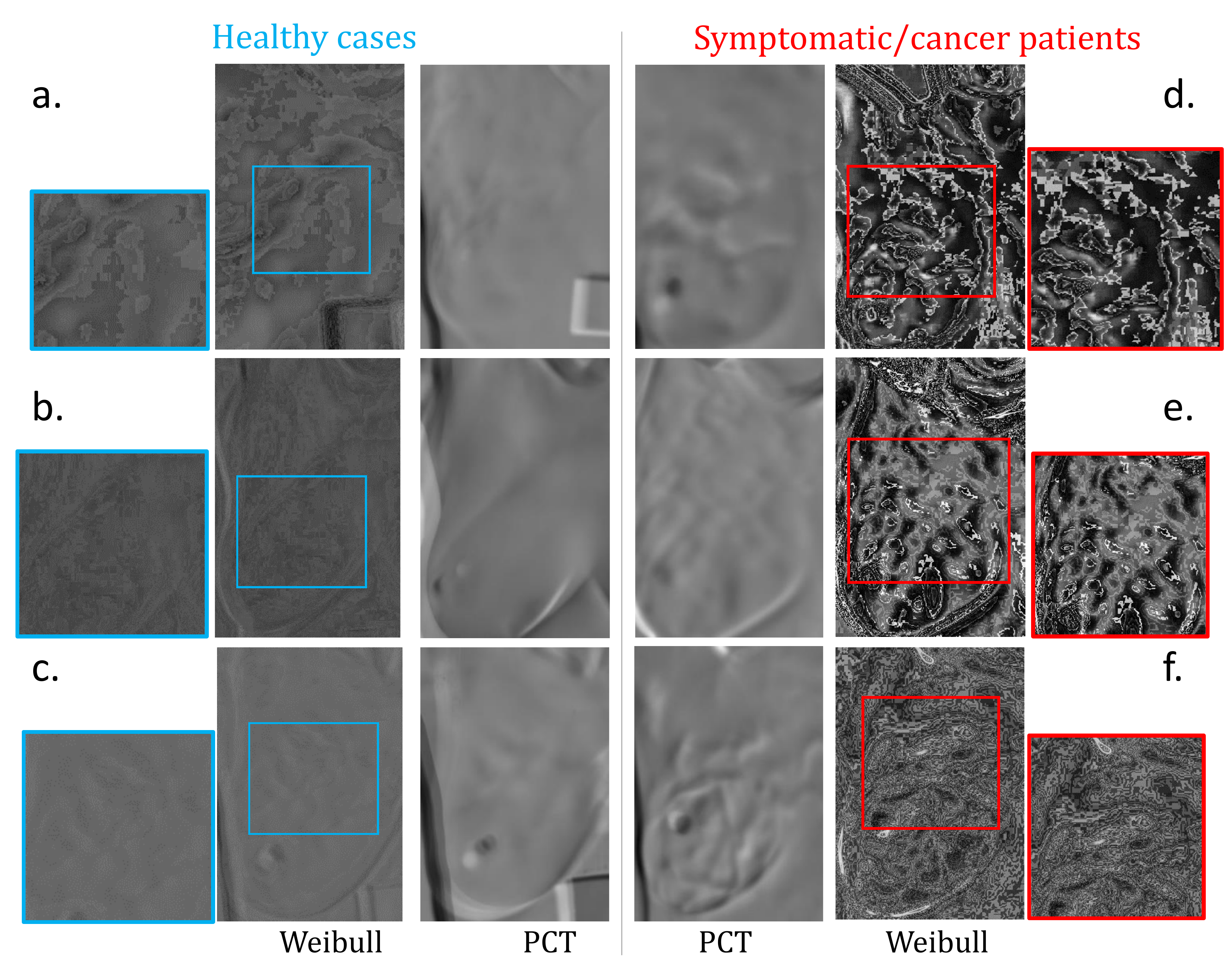}
\end{center}
   \caption{Comparing PCT to Weibull embedding thermography results, we present six instances, three from healthy cases (a-c) and three from abnormal cases (d-f). The application of Weibull embedding significantly enhances the delineation of heterogeneous thermal patterns associated with vasodilation in breast cancer cases. }
\label{fig4}
\end{figure}

\begin{figure}[t]
\begin{center}
   \includegraphics[width=0.75\linewidth]{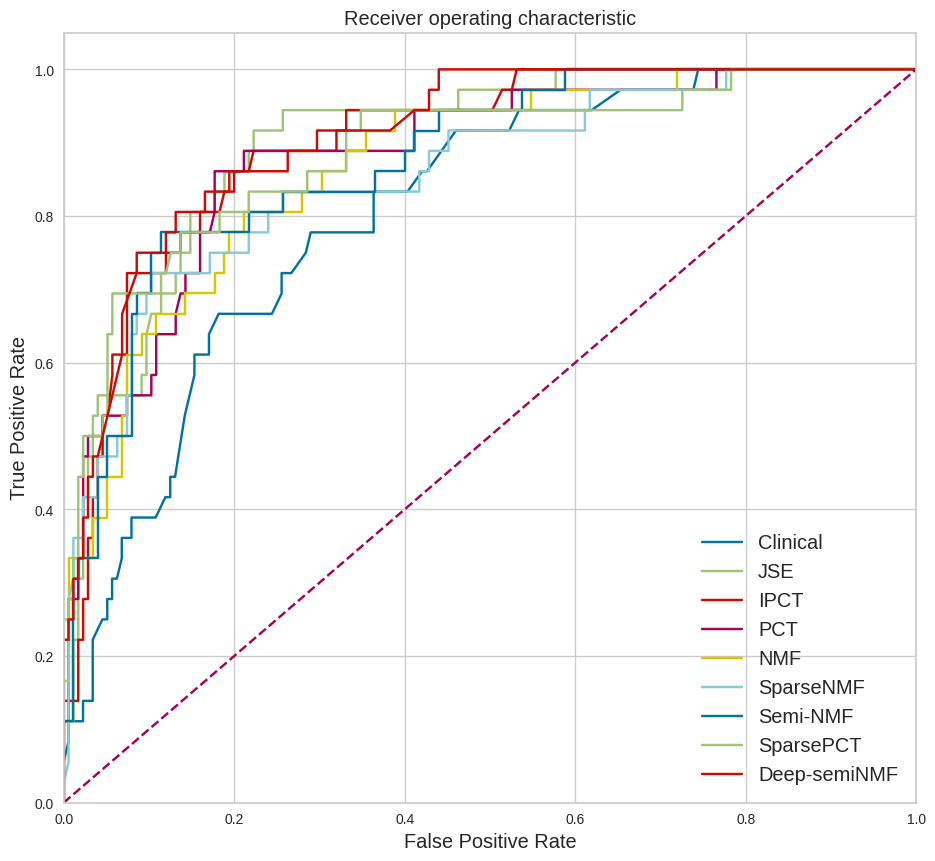}
\end{center}
   \caption{Reciever operating characteristic (ROC) curve of the model used for evaluation of Weibull-embedding thermal data for different matrix factorization techniques. }
\label{fig5}
\end{figure}

The proposed Weibull embedding approach was used to embed 5 LD-approximated basis vectors attained through different matrix factorization methods. To determine the heterogeneity level in the ROI, a thermal reference marker was attached between the participants' breasts and used to normalize the image representations. The application of the embedding technique significantly amplifies the thermal heterogeneity, enabling clear discrimination between symptomatic and cancer patients from healthy cases (Figure \ref{fig4} (\textbf{a-c})). Figure \ref{fig5}, Tables III and IV present the results of Weibull embedding in the detection cancer leading abnormalities.
Weibull embedding provided relatively higher accuracy than other stat-of-the-art embedding approaches, which tested with four different classifiers, i.e., random forest, k-nearest neighbor (KNN), Naive Bayes, and XGBoost classifiers (Table III).

\subsection{Thermomics and classification results}
A total of 354 thermomics were extracted from the breast areas, specifically the ROI, which was obtained by the embedded Convex-NMF-generated basis. These thermomics encompassed four distinct feature groups: Shape, first-order statistics, grey-level, and spatiotemporal features. The Pyradiomics python library~\cite{r32} was employed for extracting the thermomic features. Subsequently, the features were stacked into a matrix with 354 thermomics associated with every frame. To address collinearity arising from the high dimensionality attributes issue, spectral embedding was utilized to reduce the dimensionality to 7 features, which was determined to be the optimal number of thermomics.

To evaluate the initial hypothesis regarding the potential of LD embedded thermal heterogeneity as a biomarker, a random forest classifier was used to classify between healthy and unhealthy cases. The training of this model involved applying the LD thermomics generated from the study cohort. To optimize the performance of the classifier, a grid search hyperparameter tuning method was employed and optimized internal classifier's parameters were found, i.e., the maximum depth of the trees, the random state within each tree, and the number of trees in the forest.
This tuning process aimed to enhance the predictive capabilities of the model and maximize its effectiveness in identifying relevant patterns and discriminating between different classes within the dataset. 

In addition, to highlight the predictive power of Weibull embedding, we performed a comparison assessment among different low-rank approximation techniques along with the proposed embedding approaches using cross-validated random forest model. The highest accuracy was sparse PCT 80.9\% (73.6\%, 80.9\%), while the lowest accuracy was generated from Sparse NMF, 72.1\% (67.8\%, 76.2\%) with the $\kappa$-coefficients of 76.2 (72.0, 79.4) and 76.2 (71.6, 78.9), respectively.
For the subset of thermomics applying various factorization methods through Weibull embedding, a multivariate random forest classifier was trained and tested to classify patients as abnormal or healthy.

The accuracy achieved for clinical information and demographics was 73.6\% ($\pm$.5\%) and the covariates used for the clinical and demographics were family history, age, and marital status. After Sparse PCT, Deep Semi-NMF had the highest classification accuracy of 78.6\% (72.6\%, 80.9\%). PCT, NMF, and SemiNMF showed closer accuracy range of 76%


The contributions presented here were implemented in Python. JSE and Weibull embedding demonstrated lower computational time compared to other methods, contrasting with Sparse PCT and Sparse NMF.

\begin{table*}[]
\begin{center}
    
\caption{The predictive power of low-dimensional thermal patterns. } \vspace{0.1in}
\begin{tabular}{cccc}
\hline
\textbf{Binary Classifiers} & \textbf{Alpha Embedding} & \textbf{Bell-embedding} & \textbf{Weibull-Embedding} \\ \hline
Random Forest               & 78.2 (±4.1)              & 78.9 (±3.4)             & 81.7 (±5.2)                \\
KNN                         & 77.6 (±5.8)              & 75.6 (±5.1)             & 80.2 (±6)                  \\
Naïve Bayesian              & 78.2 (±4.1)              & 78.2 (±5.4)             & 79.6 (±2.8)                \\
XGBOOST                     & 64.5 (±7.1)              & 70.9 (±9.3)             & 78.1 (±12.7)               \\ \hline
\end{tabular}
\end{center}
\label{table-2}
\end{table*}

\begin{table*}[]
\begin{center}
    
\caption{The results of the cross-validated random forest classification model.} \vspace{0.1in}
\begin{tabular}{ccccc}
\hline
\multicolumn{2}{c}{\textbf{Model}}           & \textbf{Factorization method} & \textbf{\begin{tabular}[c]{@{}c@{}}Weibull embedding \\ classification accuracy $^*$ (\%)\end{tabular}} & \textbf{\begin{tabular}[c]{@{}c@{}}Kappa \\    coefficient2 ($\kappa$)\end{tabular}} \\ \hline
\parbox[t]{2mm}{\multirow{7}{*}{\rotatebox[origin=c]{90}{\textbf{Random Forest}}}} & \multicolumn{2}{c}{CCIPCT}       & 76.1 (68.9, 84.5)                                         & 77.7 (71.9, 80.0)                                                             \\
                                          & \multicolumn{2}{c}{PCT}          & 76.1 (73.6, 76.2)                                         & 75.5 (70.9, 78.9)                                                             \\
                                          & \multicolumn{2}{c}{NMF}          & 76.1 (67.9, 80.9)                                         & 75.9 (73.4, 79.4)                                                             \\
                                          & \multicolumn{2}{c}{Sparse NMF}   & 72.1 (67.8, 76.2)                                         & 76.2 (71.6, 78.9)                                                             \\
                                          & \multicolumn{2}{c}{Semi NMF}     & 76.1 (71.4, 76.1)                                         & 75.5 (71.2, 78.6)                                                             \\
                                          & \multicolumn{2}{c}{Sparse PCT}   & 80.9 (73.6, 80.9)                                         & 76.2 (72.0, 79.4)                                                             \\
                                          & \multicolumn{2}{c}{Deep SemiNMF} & 78.6 (72.6, 80.9)                                         & 78.5 (75.0, 81.1)                                                             \\ \hline
\end{tabular}
\end{center}
\label{tabel-3}

{\footnotesize
\hspace{1.0 in} $^*$ Classification accuracy reported by median ($\pm$IQR) (Interquartile range-IQR).

}
\end{table*}

\section{Discussion}
This study presents two contributions to determine the predominant and embedding of LD basis vectors for thermography. It offers an enhanced representation of thermal texture for diagnostic applications. The primary objective was to enhance the performance of LD representation for HD infrared data by modifying the LD projection of HD data and replacing conventional matrix factorization techniques with the JSE. By incorporating the JSE estimator, the proposed design sought to optimize the LD embedding process, resulting in improved LD approximation and effectiveness in capturing essential features of the thermal patterns. The Weibull embedding was also proposed and used to enhance the eigenvector's process, and a new embedding procedure was defined to unravel multiple LD representations and associate it with early diagnosis of breast cancer-leading abnormalities. The presented methods have the potential to identify possible breast cancer cases non-invasively and cost-effectively as an initial diagnostic tool with BCE before mammography or MRI.

The proposed JSE and Weibull embedding methods demonstrated enhancement in performance for detecting abnormal thermal patterns in breast thermography compared to the previously used Gaussian~\cite{r18,r19} and Bell~\cite{r22} embedding functions. JSE provided better intensity and contrast profiles of the predominant thermal basis than PCT, leading to more successful detection of heterogeneous thermal patterns. This is attributed to the correction of JSE for the leading thermal basis vector. Weibull embedding also represented a good accuracy (slightly higher) compared to other thermal basis embedding techniques. This may be due to the probability density function of Weibull random variables used to form this embedding which boosts the contrast of thermal basis higher than Gaussian~\cite{r18,r19} and Cauchy~\cite{r22} functions.

Sparse PCT and Deep SemiNMF exhibited marginally higher accuracies than other matrix factorization methods, Table IV. This is attributable to sparsity constrains in basis decomposition and multilayer basis which aggravate constrains that better highlight dominating thermal patterns that provide sparse behavior and induce stability in catching thermal patterns. CCIPCT, PCT, NMF, and SemiNMF showed relatively similar accuracy but different interquartile ranges. This may be due to the Weibull distribution that projects data to high thermal intensities leading to aggregating outcomes to reach slight saturation.

The method proposed in this study produces HD thermomics that may increase the risk of overfitting the random forest model, a consequence of the \textit{curse of dimensionality}~\cite{r33}. To address this issue, we employed spectral embedding to capture the essential features that represent the dominant thermal patterns, thereby shrinking the thermomics' dimensionality and enhancing the method's robustness. This approach aligns with other similar methods proposed previously~\cite{r18,r20,r22} and provides resilience against minimal motion artifacts initiated by the patient's movement or noise.
Nonetheless, this study has some limits related to the small cohort of patients, which restricts the statistical significance of the results. A bigger cohort of patients could highlight the importance of a dimensionality reduction method to extract reliable thermal characteristics.

This study presented JSE for estimating thermal basis and Weibull embedding in an infrared-based early breast cancer system that offers several improvements. This is the first time these two methods have been introduced for computational thermography and measuring thermal heterogeneity. Firstly, embedding the system excludes manually selecting important bases in factorizing thermal images, which is a significant aid to thermographic systems. Secondly, JSE corrected the leading thermal basis, which leads to capturing more reliable heterogeneous textures. Thirdly, the recommended approach improves the impact of motion artifacts as a significant aid to the infrared systems due to their sensitivity to the setup environment.

\section{Conclusions}
This study addressed a significant challenge in the LD approximation of thermal stream, specifically the selection of a leading basis, by applying JSE correction and introducing the Weibull embedding approach to highlight thermal heterogeneity. The effectiveness of these methods was investigated and tested on 212 thermal breast cancer screening cases. A total of 354 thermomics were extracted from the results of JSE and embedded to encode heterogeneous patterns and utilize them in the development of a diagnostic model. To evaluate the performance of the proposed approach, a comparative analysis was conducted against state-of-the-art thermographic methods, including NMF, CCIPCT, PCT, SemiMNF, Sparse PCT, Deep semi-NMF, and Sparse NMF, as well as Gaussian and Bell embedding approaches.

The qualitative and quantitative results indicated that JSE provides substantial correction to the principal thermal basis compared to PCT itself. Also, Weibull embedding demonstrated significant capability in preserving thermal heterogeneity, enabling discrimination between abnormal and healthy participants, and achieved the maximum accuracy of 81.7\% ($\pm$5.2) using convex NMF. The highest accuracy of 80.9\% (73.6\%, 80.9\%) was obtained from Sparse PCT in comparison to other factorization methods. The classification accuracy of JSE yielded 76.2\% (71.4\%,80.9\%). 

Future work will focus on further developing methods to improve thermal patterns measurement capability through more sophisticated approaches preserving the optimal predominant representation of thermographic images and enabling infrared-based diagnostic systems to work more efficiently.

\section*{ACKNOWLEDGMENTS}
The author would like to thank the respected reviewers for providing their valuable insight to improve this study. This work was supported, in part, by the University System of Maryland's William E. Kirwan Center for Academic Innovation grant 2022-2023. Thanks to Leonardo D. Buitrago for his help in performing the initial analysis.




\end{document}